\documentclass[11pt,a4paper]{article}
\pdfoutput=1 

\usepackage{authblk}
\usepackage{graphicx}
\usepackage{float}
\usepackage{subcaption}
\usepackage{fullpage}
\usepackage{amsmath}
\usepackage{amssymb}
\usepackage{multirow}
\usepackage[numbers,sort&compress]{natbib}

\newcommand{\bmat}{\left[\begin{matrix}}
\newcommand{\emat}{\end{matrix}\right]}

\title{\textbf{Steady State of Pedestrian Flow in Bottleneck Experiments}}
\author[1,2,*]{\small Weichen Liao}
\author[2]{Antoine Tordeux}
\author[2]{Armin Seyfried}
\author[2]{Mohcine Chraibi}
\author[2]{Kevin Drzycimski}
\author[3]{Xiaoping Zheng}
\author[4]{Ying Zhao}
\affil[1]{College of Information Science and Technology, Beijing University of Chemical Technology, Beijing 100029, China}
\affil[2]{J\"{u}lich Supercomputing Centre, Forschungszentrum J\"{u}lich GmbH, J\"{u}lich 52425, Germany}
\affil[3]{Department of Automation, Tsinghua University, Beijing 100084, China}
\affil[4]{Centre for Information Technology, Beijing University of Chemical Technology, Beijing 100029, China}
\date{}

\begin{document}
\maketitle

\begin{abstract}

Experiments with pedestrians could depend strongly on initial conditions. Comparisons of the results of such experiments require to distinguish carefully between transient state and steady state. In this work, a feasible algorithm - Cumulative Sum Control Chart - is proposed and improved to automatically detect steady states from density and speed time series of bottleneck experiments. The threshold of the detection parameter in the algorithm is calibrated using an autoregressive model. Comparing the detected steady states with previous manually selected ones, the modified algorithm gives more reproducible results. For the applications, three groups of bottleneck experiments are analysed and the steady states are detected. The study about pedestrian flow shows that the difference between the flows in all states and in steady state mainly depends on the ratio of pedestrian number to bottleneck width. When the ratio is higher than a critical value (approximately 115 persons/m), the flow in all states is almost identical with the flow in steady state. Thus we have more possibilities to compare the flows from different experiments, especially when the detection of steady states is difficult.

\end{abstract}


\newpage

\tableofcontents

\newpage

\section{Introduction}

In recent years, several experiments under well-controlled laboratory conditions were carried out to explore pedestrian characteristics in bottlenecks. Most experiments focused on the relationship between bottleneck width and pedestrian flow \cite{Mueller1981, Hoogendoorn2004c, Hoogendoorn2005, Nagai2006, Kretz2006a, Seyfried2009, Liddle2009, Daamen2010, Seyfried2010, Rupprecht2011, Schadschneider2011, SongW2011, TianW2012, SongW2013, LiaoW2014a, LiaoW2014b}. The flow was once announced to grow in a stepwise manner along with lane formation \cite{Hoogendoorn2005}. However, it was shown afterwards by more detailed experiments to be linearly dependent on bottleneck width. The slope of the linear function is approximately 1.9 (m$\cdot$s)$^{-1}$ with bottleneck width ranging from 0.7 m to 5.0 m \cite{Seyfried2009, Rupprecht2011, LiaoW2014a}. When bottleneck width is smaller than 0.7 m, the slope of the linear function decreases with increasing bottleneck width \cite{Kretz2006a, SongW2011, SongW2013}. Other bottleneck geometrical factors were also studied. Shorter bottlenecks provide higher flow than longer ones \cite{Liddle2009, Seyfried2010, Rupprecht2011, Schadschneider2011}. Wider width of the passage in front of bottleneck leads to higher flow \cite{Mueller1981, Rupprecht2011}. The flow also increases with increasing distance between bottleneck and holding area \cite{Rupprecht2011}. Usually bottleneck geometry was made up of boards higher than 2.0 m to prevent pedestrians' bodies overlapping the boundaries. But Nagai et al. \cite{Nagai2006} used desks with the height of approximately 0.8 m, which was pointed out to actually provide wider bottleneck widths \cite{LiaoW2014a}. Helbing et al. \cite{Helbing2005} and Yanagisawa et al. \cite{Yanagisawa2009} studied the influence of an obstacle in front of the bottleneck, and found that the existence of the obstacle leads to higher flow especially when shifted from the center. Furthermore, the influences of non-geometrical factors were studied. Nagai et al. \cite{Nagai2006} changed the initial density of the participants in the holding area, and found that the flow increases with increasing initial density but the rate of the increase decreases. Daamen et al. \cite{Daamen2010} changed the composition of the participants, and found that the experiment with mainly children has the highest flow and that with disabled pedestrians has lower flow. Most experiments were conducted under normal situations, in which the participants were asked to walk through the bottleneck with normal speed. Only a few experiments were conducted under competitive \cite{Muir1996}, hurried \cite{Helbing2003}, pushing \cite{Helbing2005} or stressful \cite{Daamen2010} situations. Unfortunately, no coincident result is made considering the influence of different situations.

It is worth noting that pedestrian movement includes transient state and steady state. In pedestrian dynamics experiments, transient state depends strongly on initial conditions while steady state is a good indicator of the independency of initial conditions, especially when the duration of experiments is short. Thus steady state is a distinguished significance for the interpretation of the results, and transient state should be excluded when combining different experiments to get universal conclusions. However, few research about bottleneck experiments considered the difference between transient state and steady state. Cepolina \cite{Cepolina2009} showed the time series of pedestrian flow. The flow changes significantly in transient state which is at the beginning and at the end, but keeps constant in steady state. Yet she did not further analyse the flow in steady state. Seyfried et al. \cite{Seyfried2009} used regression analysis to calculate the stationary values for the density and speed, but the flow in steady state was not considered. Rupprecht et al. \cite{Rupprecht2011} found that the flows with all participants and with specified participants have different trends, but the specified participants were selected arbitrarily not considering steady state. Liao et al. \cite{LiaoW2014a} compared the flows in all states and in steady state, and found that they are both linearly dependent on bottleneck width but with different slopes. Nevertheless, the duration of steady state was determined according to ``small fluctuations'' in density and speed time series of the experiments, but the demarcation of the ``small fluctuations'' is ambiguous. Moreover, the specific relationship between steady state and the flow is unknown. Up to now, no generally accepted method was used to detect steady state in bottleneck experiments. Manually selected steady state can be used, but the result varies with different steady states from different researchers. For automatical method, Krausz et al. \cite{Krausz2012} did similar work by using a Cumulative Sum Control Chart (CUSUM) algorithm to detect transitions from the optical flow computations at the Loveparade 2010 in Duisburg.

In this work, a feasible method to detect steady state is proposed and the steady state of pedestrian flow in bottleneck experiments is studied. The remainder of the paper is organized as follows. In section 2 a CUSUM algorithm is proposed and improved to automatically detect steady state, and the threshold of the detection parameter in the algorithm is calibrated using an autoregressive model. Section 3 applies the modified CUSUM algorithm to three groups of bottleneck experiments, and compares the flows in all states and in steady state. Finally, the conclusions are made in section 4.

\section{Detection of steady state}

\subsection{Relevant variables}

In pedestrian dynamics, time series of the variables density $\rho$, speed \emph{v} and flow \emph{J} and their steady states are used to characterize the transport of pedestrian streams. The classic definition of flow is the number of the pedestrians passing a line per unit time:
\begin{subequations}
    \begin{align}
        & J = \frac{N}{T} = \frac{1}{\overline{\Delta t_{i}}}, \tag{1}
    \end{align}
\end{subequations}
where $\Delta t_{i}$ is the time gap between the crossing of two following pedestrians, and \emph{T} is the time for all the pedestrians \emph{N} passing the measurement line. A reasonable resolution is necessary to identify steady state, but microscopic measurements of the flow by time gaps or mean values over small intervals could lead to strong fluctuations. These flow fluctuations are caused by the relationship between the size of the objects (here pedestrians) and the size of the facilities in combination with ordering phenomena like the zipper effect (see Figure 5 and Figure 7 in \cite{Seyfried2009}). To deal with this problem, we refer to the flow equation from fluid dynamics:
\begin{subequations}
    \begin{align}
        & J = \rho \cdot v \cdot W, \tag{2}
    \end{align}
\end{subequations}
where \emph{W} indicates the width of the facility. Here $\rho$ and \emph{v} is the mean density and mean speed in the measurement area, respectively. Following this equation, the steady states based on density and speed are a good indicator for the steady state of flow.

In this paper, we aim to detect steady state from the flow in Equation (1) with relatively short time series. The series lengths are too short to obtain precise flow estimations, so we use the density and speed variables in Equation (2) to determinate if the system is in steady state. The aforementioned density and speed are calculated using the Voronoi method to obtain estimations with low fluctuations (see Equation (8) and (9) in \cite{ZhangJ2011}). This approach allows precise detection of the steady state at the scale of only few seconds.

\subsection{CUSUM algorithm}

Since the detection of steady state is a sort of transition detection, the CUSUM algorithm is proposed in this work. It is a sequential analysis technique which is initially used for monitoring transition detection \cite{Page1954}. The precondition of the CUSUM algorithm is a reference process with observations in normal situation. Let $(x_{i})_{i=1}^{n}$ denote a sequence of observations in density or speed. The manually selected steady state can be regarded as the reference process $(x_{i})_{i=j}^{m}$, where $j \geq 1$ and $m \leq n$ (see the interval between the two green dash-dotted lines in Figure 1). Let \emph{Q($\alpha$)} and \emph{Q(1-$\alpha$)} represent the upper and lower $\alpha$-percentile of the distribution in the reference process $(x_{i})_{i=j}^{m}$, respectively. The CUSUM algorithm continuously accumulates the deviations of the observations $(x_{i})_{i=1}^{n}$ from the $\alpha$-percentiles \cite{Page1954}:
\begin{subequations}
    \begin{align}
        & s_{i}^{+} = \max\{ 0, s_{i-1}^{+} + x_{i} - Q(\alpha) \}, & s_{0}^{+} = 0; \tag{3}\\
        & s_{i}^{-} = \max\{ 0, s_{i-1}^{-} + Q(1-\alpha) - x_{i} \}, & s_{0}^{-} = 0. \tag{4}
    \end{align}
\end{subequations}
The statistics $s_{i}^{+}$ and $s_{i}^{-}$ show the degree of the fluctuations in $(x_{i})_{i=1}^{n}$. If the distribution of $(x_{i})_{i=1}^{n}$ is different from the distribution of $(x_{i})_{i=j}^{m}$, then the statistics increase. Therefore, the CUSUM algorithm is able to detect the transitions of the observations from the given reference process. The specific way is combining $s_{i}^{+}$ and $s_{i}^{-}$ with a threshold of the detection parameter $\theta$, then the interval with both $s_{i}^{+}$ and $s_{i}^{-}$ under $\theta$ is the steady state.

\begin{figure}[t]
    \centering
    \begin{subfigure}{0.49\textwidth}
        \includegraphics[height=0.32\textheight]{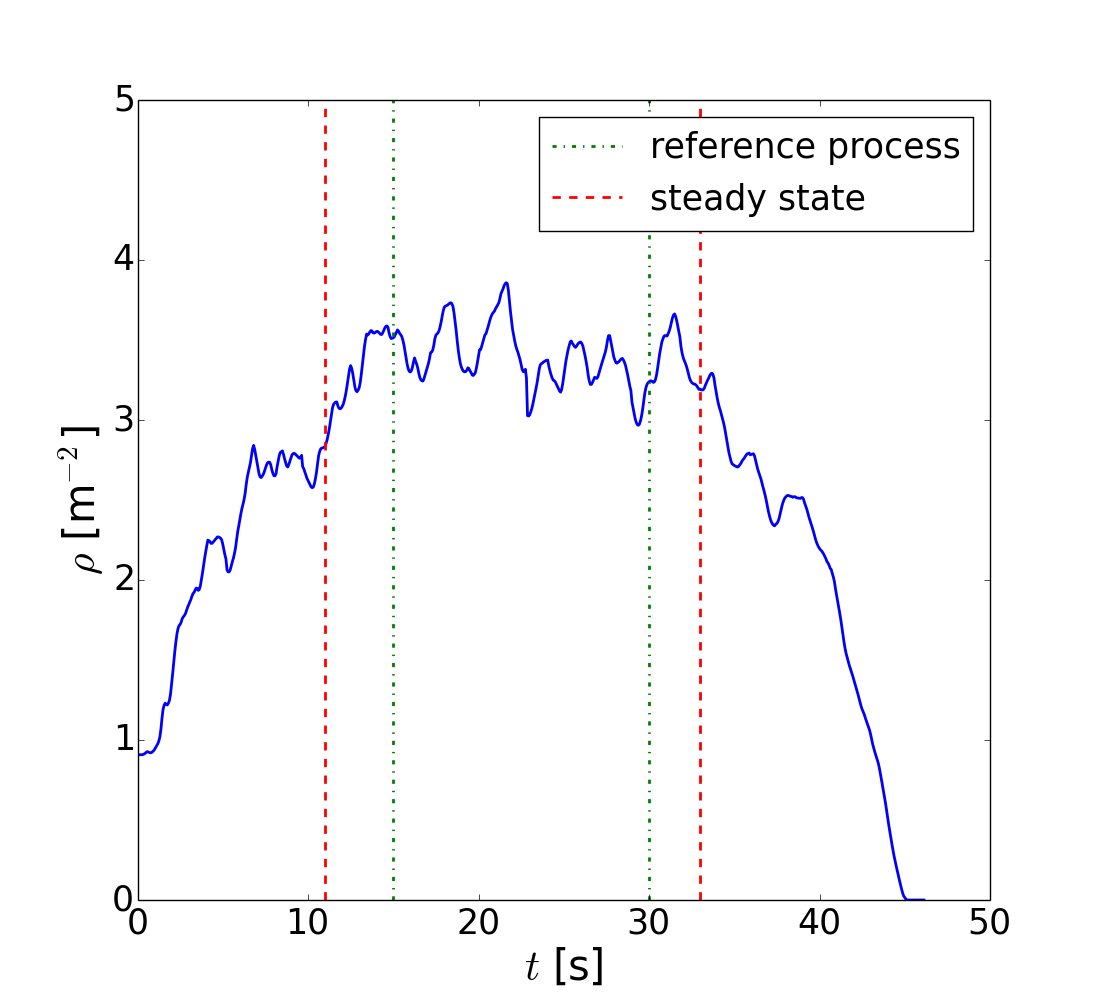}
        \caption{density}
    \end{subfigure}
    \centering
    \begin{subfigure}{0.49\textwidth}
        \includegraphics[height=0.32\textheight]{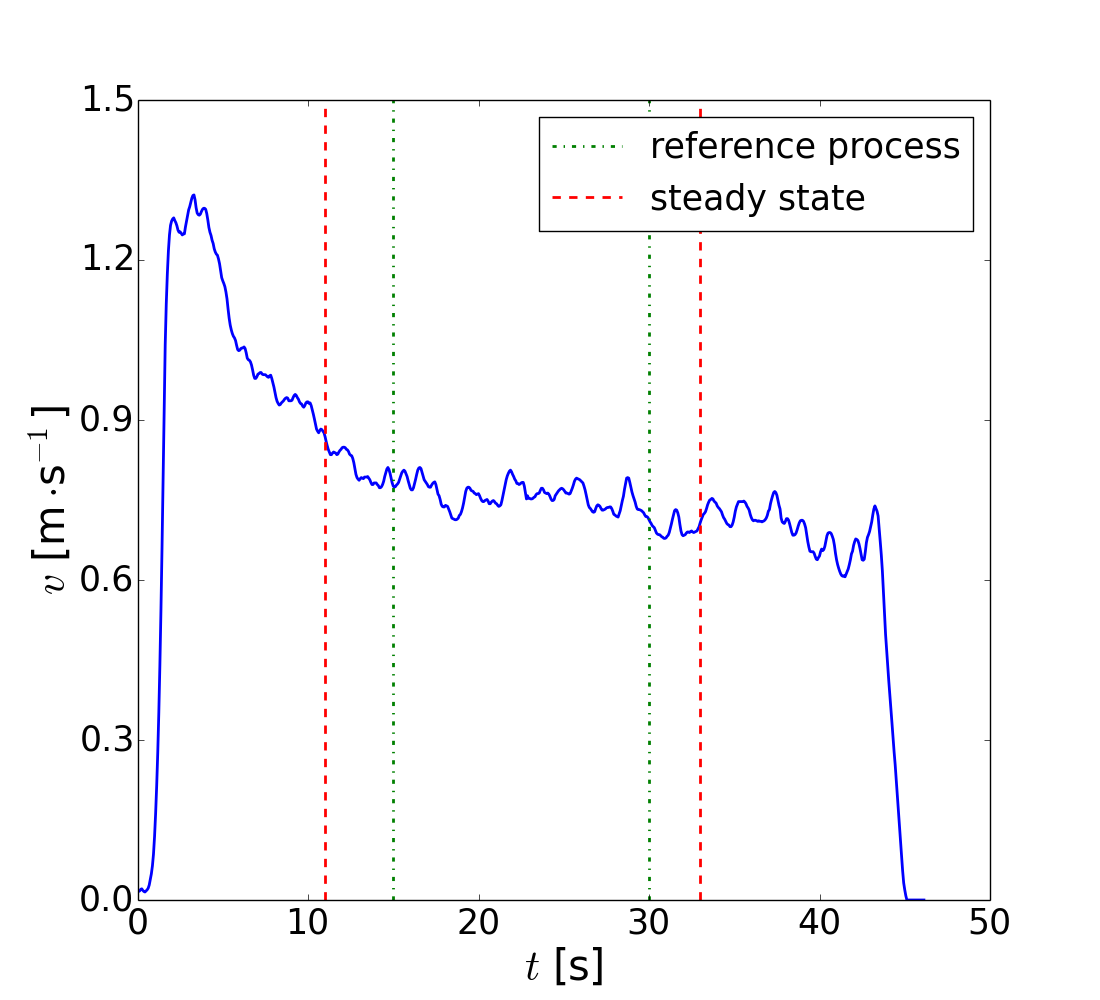}
        \caption{speed}
    \end{subfigure}
    \caption{Time series of the observations $(x_{i})_{i=1}^{n}$ in (a) density and (b) speed. The interval between the two green dash-dotted lines is the manually selected steady state, which is regarded as the reference process $(x_{i})_{i=j}^{m}$, where $j \geq 1$ and $m \leq n$. The interval between the two red dashed lines is the final steady state, which is obtained using the detected steady state minus the corresponding reaction time.}
\end{figure}

The preliminary CUSUM algorithm can detect the steady state, but there exist problems during the detection procedure. First, since the statistics $s_{i}^{+}$ and $s_{i}^{-}$ are combined to detect the steady state, it is redundant to calculate them separately. Therefore, new statistics $(s_{i})_{i=1}^{n}$ are introduced (Equation (5)). Second, the response of the statistics to the fluctuations in $(x_{i})_{i=1}^{n}$ is not sensitive enough. The increase rate and decrease rate of the statistics are not the same. To solve this problem, a step function is introduced to enhance the sensitivity of the response (Equation (6)). Third, the criterion in Equation (6) is not the $\alpha$-percentiles of $(x_{i})_{i=j}^{m}$ any more. The replacement is $q(\alpha)$, which is the upper $\alpha$-percentile in a standard normal distribution $N(0,1)$. Accordingly, the standard score transformation of the observations $(x_{i})_{i=1}^{n}$ is used (Equation (7)). Fourth, the maximum value of the statistics is very large that it needs a long time to get back to steady state again. As an improvement, a boundary $s_{\max}$ is introduced to limit the increase of the statistics. Last but not least, since it is unknown whether at the beginning the statistics are in steady state or not, the value of $s_{0}$ should be equal to the boundary $s_{\max}$. From the above, the modified CUSUM algorithm is as follows:
\begin{subequations}
    \begin{align}
    & s_{i} = \min\{ \max\{ 0, s_{i-1} + F(\tilde{x}_{i}) \}, s_{\max} \}, & s_{0} = s_{\max}; \tag{5} \\
    & F(\tilde{x}_{i}) = \tag{6}
        \begin{cases}
            1  & \text{if} \quad |\tilde{x}_{i}| > q(\alpha), \\
            -1 & \text{if} \quad |\tilde{x}_{i}| \leq q(\alpha); \\
        \end{cases}
    & \\
    & \tilde{x}_{i} = \frac{x_{i} - \mu}{\sigma}; \tag{7}
    \end{align}
\end{subequations}
where $\mu$ is the mean of $(x_{i})_{i=j}^{m}$, and $\sigma$ is the standard deviation of $(x_{i})_{i=j}^{m}$. As shown in Figure 2, the statistics $(s_{i})_{i=1}^{n}$ calculated by the modified CUSUM algorithm have sensitive response to the fluctuations, and the trend of the statistics is clear.

\begin{figure}[t]
    \centering
    \begin{subfigure}{0.49\textwidth}
        \includegraphics[height=0.32\textheight]{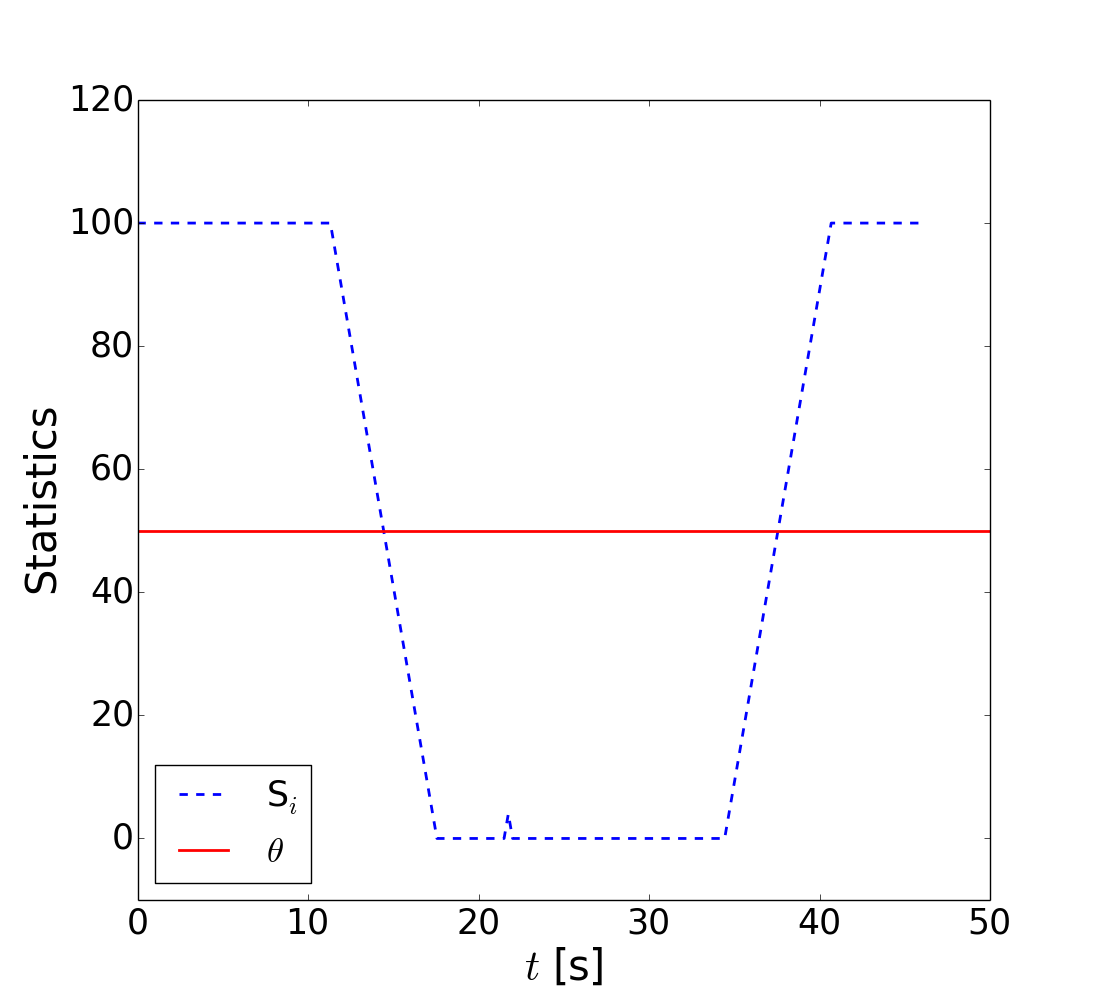}
        \caption{density}
    \end{subfigure}
    \centering
    \begin{subfigure}{0.49\textwidth}
        \includegraphics[height=0.32\textheight]{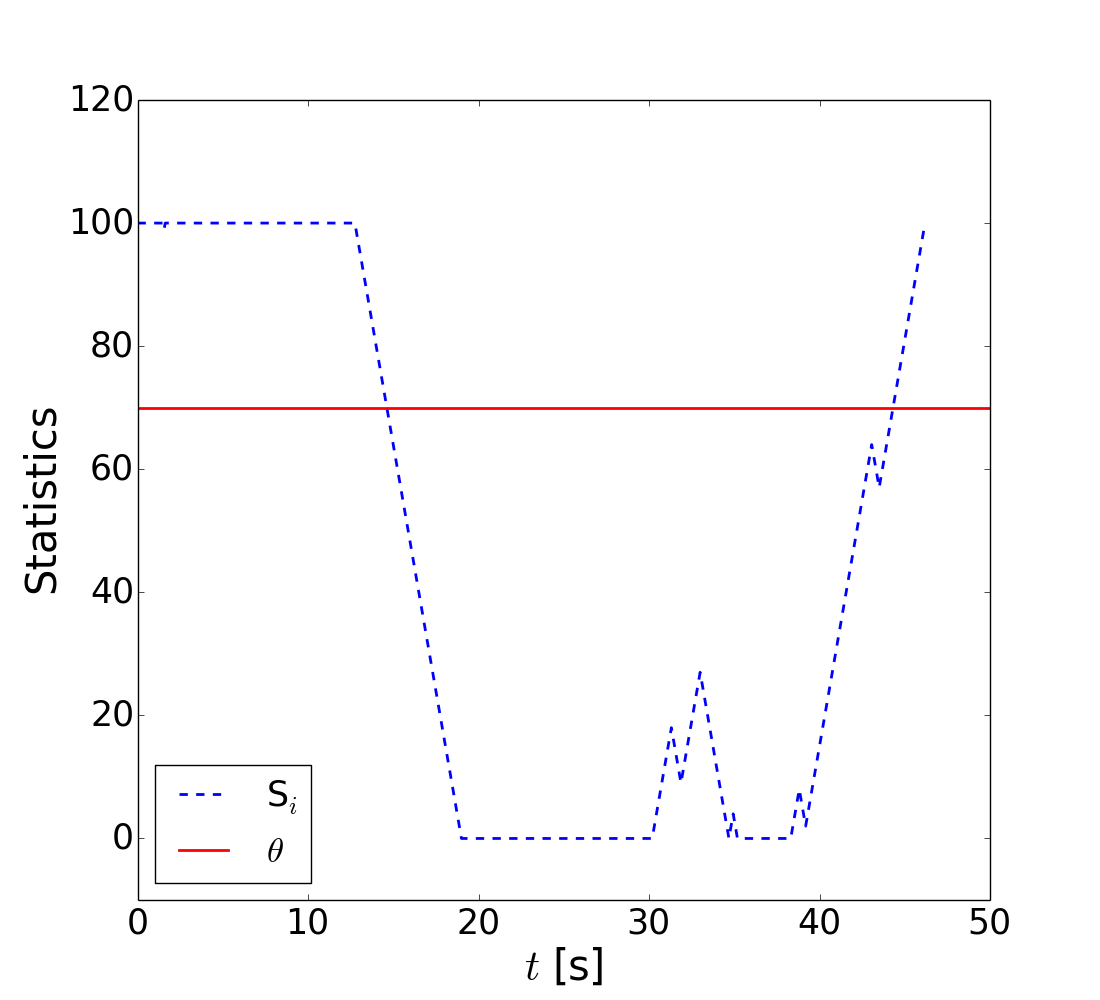}
        \caption{speed}
    \end{subfigure}
    \caption{Detection of steady state by the modified CUSUM algorithm. The statistics in (a) density and (b) speed are calculated based on the observations $(x_{i})_{i=1}^{n}$ in Figure 1 (a) and (b), respectively. $s_{i}$ shows the degree of the fluctuations in the observations $(x_{i})_{i=1}^{n}$. The interval with $s_{i}$ under $\theta$ is the detected steady state.}
\end{figure}

We should note that the detected transitions (the interactions in Figure 2) are not the real transitions in the observations $(x_{i})_{i=1}^{n}$. There is a reaction time caused by the detection procedure itself. When reaching steady state and leaving steady state, the reaction time is calculated as follows, respectively:
\begin{subequations}
    \begin{align}
        & t_{\text{reaching}} = \frac{s_{\max} - \theta}{f}, & t_{\text{leaving}} = \frac{\theta}{f}, \tag{8}
    \end{align}
\end{subequations}
where $f$ is the frame number per second in the observations. The final steady state is using the detected transitions minus the corresponding reaction times (see the interval between the two red dashed lines in Figure 1).

\subsection{Threshold of detection parameter}

The accurate detection of steady state depends on the detection parameter's threshold $\theta$ to be calibrated. For a Neyman-Pearson statistical test, $\theta(\gamma)$ is the upper $\gamma$-percentile of the statistics $(s_{i})_{i=j}^{m}$ for the reference process $(x_{i})_{i=j}^{m}$. In this way, we are able to control the probability of detecting a false transition, which is the risk level 1-$\gamma$.

Different methods to calibrate $\theta$ are shown in Figure 3. The simplest way is the direct calibration with the reference process $(x_{i})_{i=j}^{m}$ ($j \geq 1$ and $m \leq n$). This method requires that the number of the selected observations in the reference process should be large enough for calibration. If not, a bootstrap method is used to expand the reference process \cite{Krausz2012}. The requirement is that the observations in $(x_{i})_{i=j}^{m}$ should be independent of each other, or that the number of the observations in $(x_{i})_{i=j}^{m}$ should be large enough (moving block bootstrap).

\begin{figure}[t]
    \centering
    \includegraphics[height=0.3\textheight]{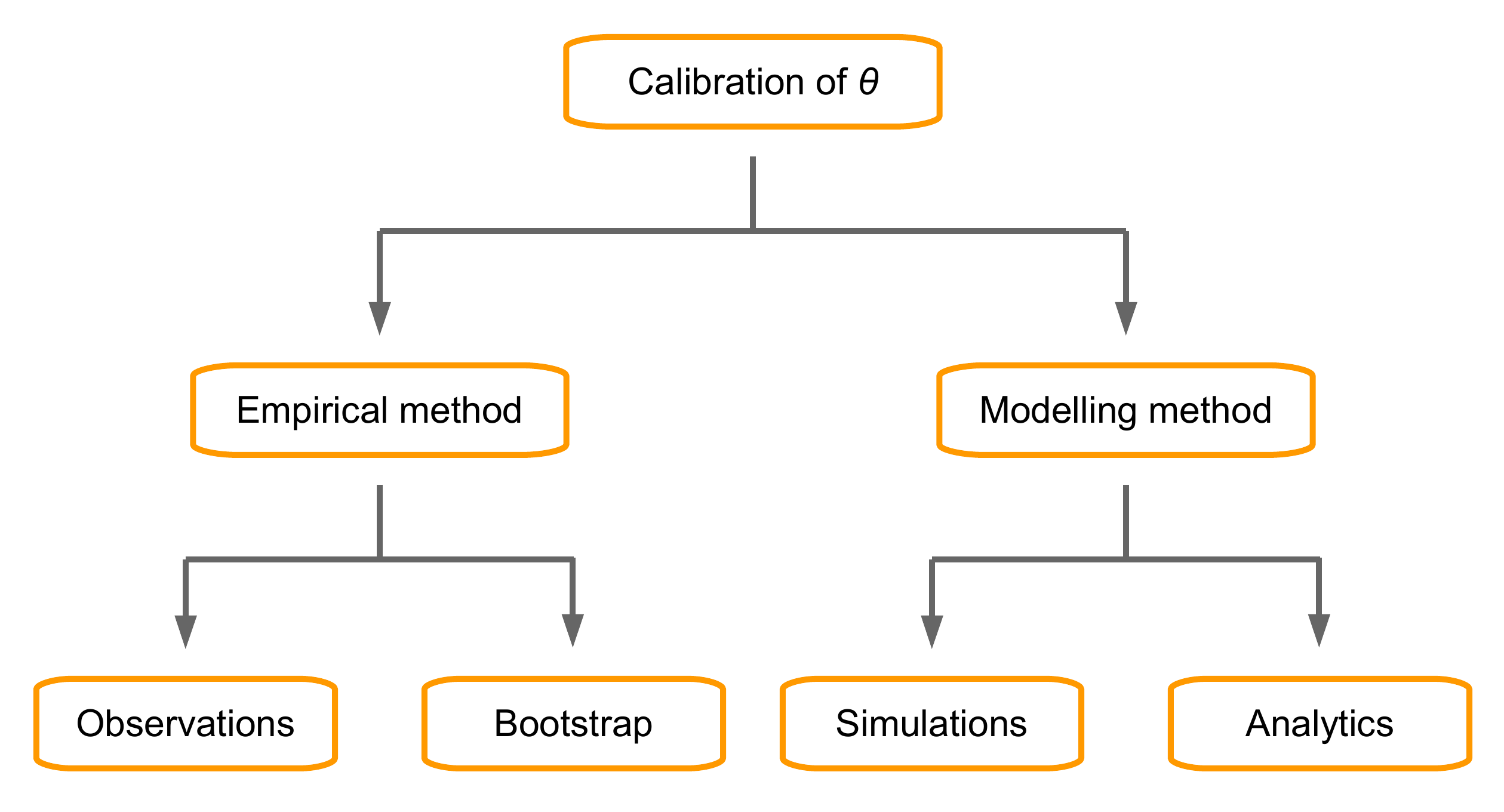}
    \caption{Different methods to calibrate the threshold of the detection parameter $\theta$.}
\end{figure}

In this work, the observations do not meet any requirements of the two empirical methods in Figure 3. Therefore, an autoregressive process $(y_{i})_{i=1}^{T}$ is proposed to model the standard score of the reference process $(\tilde{x}_{i})_{i=j}^{m}$:
\begin{subequations}
    \begin{align}
    & y_{i} = c \cdot y_{i-1} + \sqrt{ 1 - c^2 } \cdot \varepsilon_{i}, & y_{0} = 0, \tag{9}
    \end{align}
\end{subequations}
where $c$ is the first autocorrelation of $(\tilde{x}_{i})_{i=j}^{m}$, and $(\varepsilon_{i})_{i=1}^{T}$ are independent normal random variables. The autoregressive process $(y_{i})_{i=1}^{T}$ is stationary with normal distribution, and $\text{cor}(y_{i}, y_{i+1}) = c$ for all $i$. Using the autoregressive model, the threshold $\theta(\gamma)$ can be obtained by simulations or analytically by using mathematical tools of stochastic processes. If by simulations, the simulation time $T$ should be sufficiently long enough to exclude fluctuations and precisely estimate the distribution of $(s_{i})_{i=j}^{m}$. Moreover, several runs must be done to control the precision of the estimation. When $c,\gamma \approx 1$, the computational effort to get good estimations is important for standard computers. For instance if $c = \gamma = 0.99$, the process has to be simulated $T = 1\text{e}8$ steps to obtain precise results. The threshold $\theta(\gamma)$ with the autoregressive process can be faster obtained analytically. The couple $(y_{i}, s_{i})$ is a Markov chain with stationary distribution $\mu(x,s) \, \text{d} x$ ($x \in \mathbb R$, $0 \le s \le s_{\max}$) such that (global balance equation):
\begin{subequations}
    \begin{align}
    \mu(x,s) = \tag{10}
        \begin{cases}
            \int \mu(y,s-1) \, g(y,x) \, \text{d}y & \text{if} \quad |x| > q(\alpha), \\
            \int \mu(y,s+1) \, g(y,x) \, \text{d}y & \text{if} \quad |x| \le q(\alpha), \\
        \end{cases}
    \end{align}
\end{subequations}
for all $0 < s < s_{\max}$, and at the borders
\begin{subequations}
    \begin{align}
    \mu(x,s_{\max}) = \int ( \mu(y, s_{\text{max}} - 1) + \mu(y, s_{\text{max}}) ) \, g(y,x) \, \text{d}y \quad \text{and} \quad \mu(x,0) = 0, \tag{11}
    \end{align}
\end{subequations}
if $|x| > q(\alpha)$, and
\begin{subequations}
    \begin{align}
    \mu(x,0) = \int (\mu(y,1)+\mu(y,0)) \, g(y,x) \, \text{d}y \quad \text{and} \quad \mu(x,s_{\max}) = 0, \tag{12}
    \end{align}
\end{subequations}
if $|x| \le q(\alpha)$.
Here $g(y,x) = \frac{1}{\sqrt{2 \pi (1-c^2)}}\exp\left(-\frac{(x-c y)^2}{2 (1-c^2)}\right)$.

Unfortunately, Equations (10-12) do not have an explicit solution for $\mu(\cdot\,,\cdot)$. Yet they can be approximated by using the rectangular numerical scheme $x_{i} = -x' + \delta_K i$ with $\delta_K = 2 x' / K$, $a_{i,s} = \mu(x_i,s)$ and $b_{i,k} = \delta_K\, g(x_k,x_i)$. The numerical approximation yields in the linear equation
\begin{subequations}
    \begin{align}
    M \mathbf x = 0, \tag{13}
    \end{align}
\end{subequations}
where
\begin{subequations}
    \begin{align}
\mathbf x = \, ^T\!\!\left(a_{0,0}, a_{1,0}, \ldots a_{K,0}, a_{0,1}, \ldots a_{K,s_{\max}}\right), \tag{14}
    \end{align}
\end{subequations}
and $M = B - \text{Id}$ with
\begin{subequations}
    \begin{align}
    B =\!\bmat
    B_2&B_2&&&\\
    B_1&&B_2&&\\[5mm]
    &&B_1&&B_2\\
    &&&B_1&B_1\emat,\quad
    B_1 =\!\bmat
    b_{0,0}\ldots\  b_{0,K}\\[2mm]
    b_{i_a,0}\ldots\  b_{i_a,K}\\[2mm]
    \\[-0mm]
    b_{i_b,0}\ldots\  b_{i_b,K}\\[2mm]
    b_{K,0}\ldots\  b_{K,K}
    \emat,\quad
    B_2 =\!\bmat
    \\[2mm]
    b_{i_a+1,0}\ldots\  b_{i_a+1,K}\\[2mm]
    b_{i_b-1,0}\ldots\  b_{i_b-1,K}\\[2mm]
    \
    \emat, \tag{15}
    \end{align}
\end{subequations}
$i_a = \text{arg} \max_i\{x_i < -q(\alpha)\}\;$ and $\;i_b = \text{arg} \min_i\{x_i > q(\alpha)\}$.
The stationary distribution of the couple $(y_i,s_i)$ is approximated by solving a linear system with $(K+1)(s_{\max}+1)$ equations. The complexity of the resolution is in $O(s_{\max}K^3)$ by using the Thomas-Algorithm (see in Appendix), which makes in general the numerical approximation of the analytical solution faster than the simulations. It is difficult to estimate the error of the numerical approximation.
Tests show that $K \approx 50$ gives good results. The error can be reduced by using trapezoidal or polynomial schemes instead of rectangular ones.

\subsection{Robustness}

For a series of observations $(x_{i})_{i=1}^{n}$, the steady state detected by the modified CUSUM algorithm depends on the combination of the statistics $(s_{i})_{i=1}^{n}$ and the threshold of the detection parameter $\theta$. Scrutinizing Equations (5-7) and (9), six key parameters determining the values of $(s_{i})_{i=1}^{n}$ and $\theta$ are summarized in Table 1. $\alpha$, $\gamma$ and $s_{\max}$ are independent of the observations so they should be calibrated independently. The value of $\alpha$ should be close to 1 to diminish outliers in $(x_{i})_{i=j}^{m}$ as much as possible. The value of $\gamma$ should be close to 1 to diminish false detections as much as possible. The boundary $s_{\max}$ of the statistics $(s_{i})_{i=1}^{n}$ should be approximately two times of $\theta$. In this work, the values of the three calibrated parameters are fixed, and $\alpha$ = 0.99, $\gamma$ = 0.99, $s_{\max}$ = 100. On the contrary, $\mu$, $\sigma$ and $c$ vary with different observations because they are measured from the original reference process $(x_{i})_{i=j}^{m}$. $\mu$ and $\sigma$ affect the statistics $(s_{i})_{i=1}^{n}$ (Equation (7)), and $c$ affects the threshold of the detection parameter $\theta$ (Equation (9)). In the situation that $(s_{i})_{i=1}^{n}$ and $\theta$ both change with different reference processes, the detected steady states might be different.

\begin{table}[t]
    \caption{Key parameters in the modified CUSUM algorithm.}
        \begin{center}
            \begin{tabular}{|c|c|l|}
                \hline
                \multirow{3}{*}{Calibrated parameters} & $\alpha$ & critical probability for the reference distribution $(x_{i})_{i=j}^{m}$ \\
                 & $\gamma$ & confident level \\
                 & $s_{\max}$ & boundary of the statistics $(s_{i})_{i=1}^{n}$ \\
                \hline
                \multirow{3}{*}{Measured parameters} & $\mu$ & mean of the observations in $(x_{i})_{i=j}^{m}$ \\
                 & $\sigma$ & standard deviation of the observations in $(x_{i})_{i=j}^{m}$ \\
                 & $c$ & autocorrelation of the observations in $(x_{i})_{i=j}^{m}$ \\
                \hline
        \end{tabular}
    \end{center}
\end{table}

To investigate the robustness of the modified CUSUM algorithm, three reference processes are manually selected from the same observations and they do not overlap with each other (see Figure 4). Keeping all the conditions the same, the detection procedure is repeated for each reference process. The results show that the three detected steady states in Figure 4 are almost the same. Moreover, they coincide with the steady state in Figure 1 (a). Therefore, for a given series of observations, the detected steady state is fixed with any reasonable reference process from the observations. By using the modified CUSUM algorithm, researchers are able to reproduce the same steady state with different manually selected reference processes.

\begin{figure}[t]
    \centering
    \includegraphics[height=0.4\textheight]{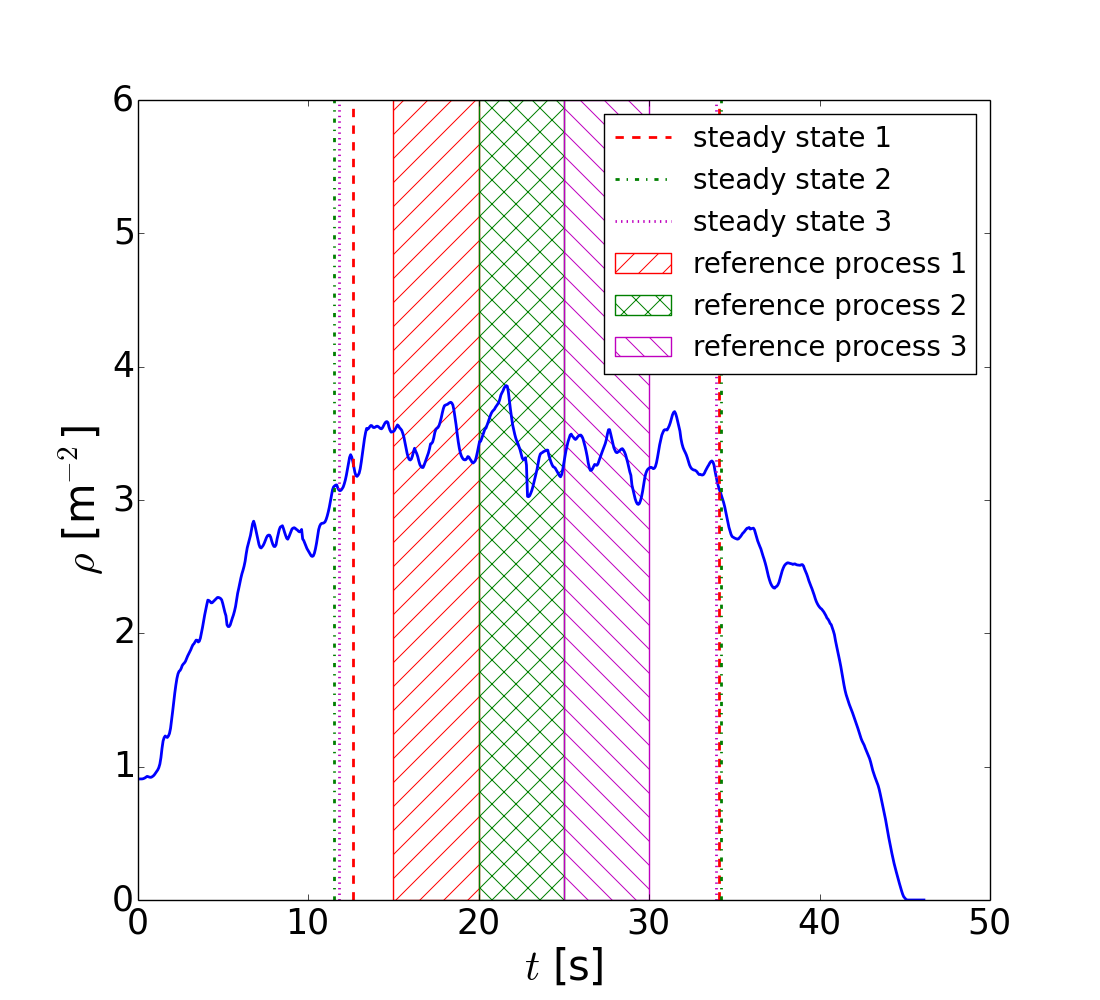}
    \caption{Time series of the observations $(x_{i})_{i=1}^{n}$ in density. The red slashed, green latticed and magenta back-slashed areas represent the manually selected reference processes. The corresponding detected steady state is the interval between the two red dashed lines, the two green dash-dotted lines and the two magenta dotted lines, respectively.}
\end{figure}

\section{Pedestrian flow in bottleneck experiments}

\subsection{Experimental setup}

Three groups of bottleneck experiments are analysed in this work. The principle setup is shown in Figure 5. Bottleneck width and bottleneck length is \emph{b} and \emph{l}, respectively. The width of the passage in front of the bottleneck is \emph{w}. The distance between the bottleneck and the holding area is \emph{d}. Before each run, \emph{N} participants are arranged in the holding area with initial density $\rho_{\text{ini}}$. The corresponding parameters in each experiment are listed in Table 2. The experiment \emph{EG} was conducted in 2006 in D\"usseldorf, and the experiments \emph{AO} and \emph{UO} were conducted in 2009 in D\"usseldorf. More information of the experiment \emph{EG}, \emph{AO} and \emph{UO} is given in \cite{Rupprecht2011}, \cite{LiaoW2014a, LiaoW2014b} and \cite{ZhangJ2011}, respectively. Note that the shape of the holding area in the experiment \emph{AO} was not rectangular but semi-circular with radius 8.6 m directly in front of the bottleneck (see Figure 1 in \cite{LiaoW2014a}). Also note that \emph{d} in the experiment \emph{UO} was 12.0 m, which included a free region of 4.0 m long and a passage of 8.0 m long (see Figure 1 (a) and (b) in \cite{ZhangJ2011}).

\begin{figure}[t]
    \centering
    \includegraphics[height=0.4\textheight]{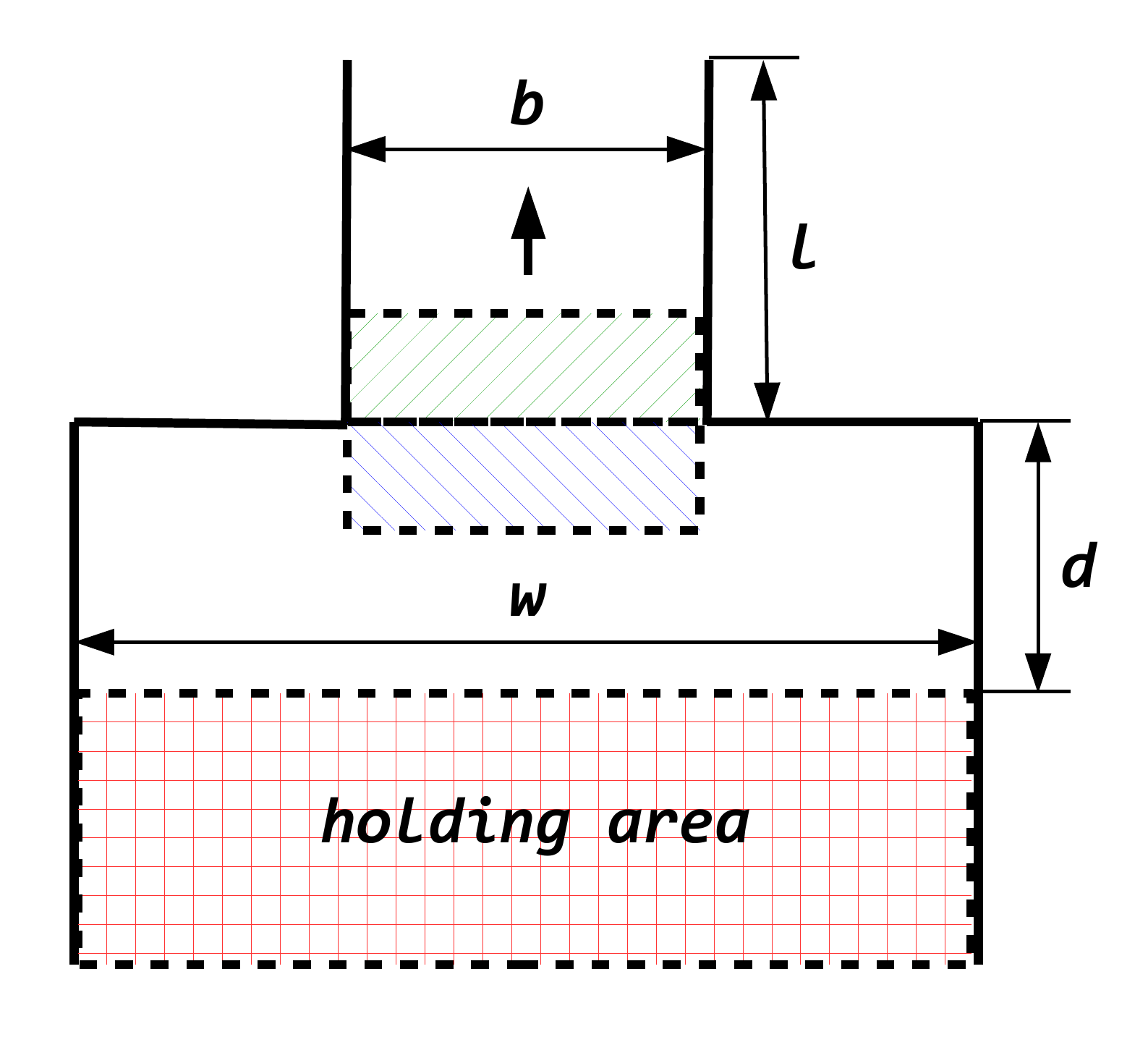}
    \caption{Experimental setup. \emph{b} is bottleneck width. \emph{l} is bottleneck length. \emph{w} is the width of the passage in front of the bottleneck. \emph{d} is the distance between the bottleneck and the holding area. Before each run, \emph{N} participants are arranged in the holding area with initial density $\rho_{\text{ini}}$. The measurement area of density and speed for the experiments \emph{EG} and \emph{AO} is slashed in green, and that for the experiment \emph{UO} is back-slashed in blue. The size of the measurement areas is \emph{b*1}.}
\end{figure}

\begin{table}[t]
    \caption{Parameters in three different bottleneck experiments \emph{EG}, \emph{AO} and \emph{UO}. \emph{b} is bottleneck width. \emph{l} is bottleneck length. \emph{w} is the width of the passage in front of the bottleneck. \emph{d} is the distance between the bottleneck and the holding area. \emph{N} is the number of participants. $\rho_{\text{ini}}$ is the initial density of the participants in the holding area.}
        \begin{center}
            \begin{tabular}{|c|c|c|c|c|c|c|c|}
                \hline
                Experiment & \emph{EG} & \emph{AO} & \emph{UO} \\
                \hline
                \multirow{3}{*}{\emph{b} [m]} & 0.9, 1.0, 1.1, 1.2, 1.4 & 2.4, 3.0, 3.6, 4.4, 5.0 & 0.7, 0.95, 1.2 \\
                 & 1.6, 1.8, 2.0, 2.2, 2.5 & & 1.0, 1.3, 1.6 \\
                 & & & 0.8, 1.2, 1.6, 2.0 \\
                \hline
                \emph{l} [m] & 4.0 & 1.0 & 0.1 \\
                \hline
                \emph{w} [m] & 7.0 & 18.0 & 1.8, 2.4, 3.0 \\
                \hline
                \emph{d} [m] & 4.0 & 0 & 12.0 \\
                \hline
                \emph{N} [persons] & 180 & 350 & 150, 250, 400 \\
                \hline
                $\rho_{\text{ini}}$ [m$^{-2}$] & 2.6 & 3.0 & 3.0 \\
                \hline
                \multirow{3}{*}{\emph{N/b} [persons/m]} & 200, 180, 164, 150, 129 & 146, 117, 97, 80, 70 & 214, 158, 125 \\
                 & 113, 100, 90, 82, 72 & & 250, 192, 156 \\
                 & & & 500, 333, 250, 200 \\
                \hline
        \end{tabular}
    \end{center}
\end{table}

The modified CUSUM algorithm is applied to detect the steady states of the three groups of bottleneck experiments. For the experiments \emph{EG} and \emph{AO} the measurement area of the detected variables, density and speed, is at the beginning of the bottleneck with size \emph{b} (see the region slashed in green in Figure 5). For the experiment \emph{UO} the measurement area is in front of the bottleneck with size \emph{b} (see the region back-shaded in blue in Figure 5). The measurement line of pedestrian flow is always at the beginning of the bottleneck.

\subsection{Steady state and flow}

The relationship between flow and bottleneck width is shown in Figure 6. All the flows are calculated according to Equation (2). The error bars represent the standard errors of the flows in steady states. The calculation of the standard error is by dividing the standard deviation by the square root of the number of the measurements that make up the mean. Since the observation (flow) in each frame is dependent with the observations nearby, they cannot be used as the measurements directly. Thus the observations in steady state are divided into groups that the number of the groups \emph{P} is the number of the measurements to calculate the standard error:
\begin{subequations}
    \begin{align}
    & P = \frac{F}{p}, \tag{16}
    \end{align}
\end{subequations}
where \emph{F} is the number of the observations in steady state, and \emph{p} is the number of the observations in each group. Here \emph{p} should be large enough to make the groups independent of each other, but also small enough to ensure the sufficient number of the measurements to diminish fluctuations when making up the mean. In this work, $p = f$ and if $P < 20$ the bias of an estimator is used. As shown in Figure 6, the flow is linearly dependent on bottleneck width but the slope of the linear function varies in different experiments. This reconfirms the result that the flow per unit width is constant as bottleneck width changes \cite{Seyfried2009, LiaoW2014a}.

\begin{figure}[t]
    \centering
    \includegraphics[height=0.4\textheight]{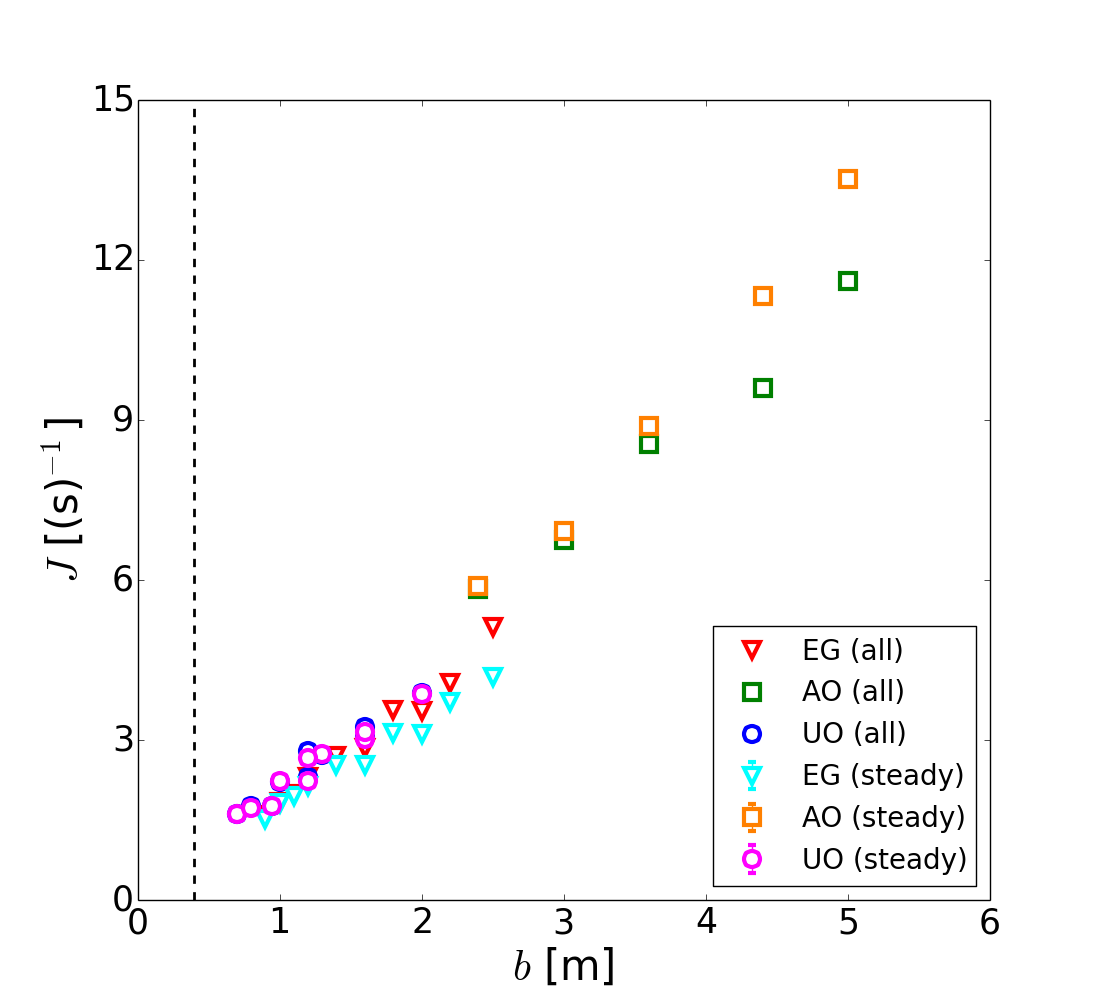}
    \caption{Relationship between flow \emph{J} and bottleneck width \emph{b}. For each experiment, the flows in all states and in steady state are represented by same symbol with different colours. The error bars represent the standard errors of the flows in steady states. The black dashed line corresponds to \emph{b} = 0.4 m, which is the minimum width for pedestrians to pass.}
\end{figure}

For each run in each experiment, the flows in all states and in steady state are compared and the difference between them is analysed. As shown in Figure 6, the difference changes significantly in the last few runs of the experiments \emph{EG} and \emph{AO} but slightly in the experiment \emph{UO}. To scrutinize what exactly impacts the difference, the parameters in Table 2 are studied. Bottleneck length \emph{l}, passage width \emph{w}, the distance between bottleneck and holding area \emph{d}, pedestrian number \emph{N} and initial density $\rho_{\text{ini}}$ are same in the experiments \emph{EG} and \emph{AO}. Since the difference changes significantly in part runs of these two experiments, the above parameters have no direct influence on the difference. Bottleneck width \emph{b} is positively correlative with the difference in the experiments \emph{EG} and \emph{AO}, but the relationship is ambiguous in the experiment \emph{UO}. In this situation, we cannot judge that bottleneck width \emph{b} is directly proportional to the difference. Afterwards, different combinations of the parameters in Table 2 are studied. The ratio of pedestrian number to bottleneck width \emph{N/b} is found to have a direct influence on the aforementioned difference, which is represented by \emph{Z} in Figure 7. \emph{Z} is calculated by using the flow in all states minus the flow in steady state and taking the absolute value. As shown in Figure 7, \emph{Z} decreases with the increase of the ratio \emph{N/b}. The critical value of the ratio \emph{N/b} is approximately 115 persons/m, which determines if the difference between the flows in all states and in steady state is acceptable. Considering the practical significance of \emph{N/b}, it describes whether the number of participants is large enough for the corresponding bottleneck width to reach a steady state. When the value of \emph{N/b} is higher than the critical value, the steady state is quickly reached and the duration of steady state is long. Conversely, it takes a long time to reach steady state and the duration of steady state is short when the value of \emph{N/b} is lower than the critical value. In the case of extremely small values of \emph{N/b}, the steady state even cannot be reached. From the above, the ratio of pedestrian number to bottleneck width \emph{N/b} is the main influence factor of the difference between the flows in all states and in steady state. Other factors, such as the composition of participants and the motivation of the experiments, also impact the difference aforementioned, but they are beyond the consideration of this work.

\begin{figure}[t]
    \centering
    \includegraphics[height=0.4\textheight]{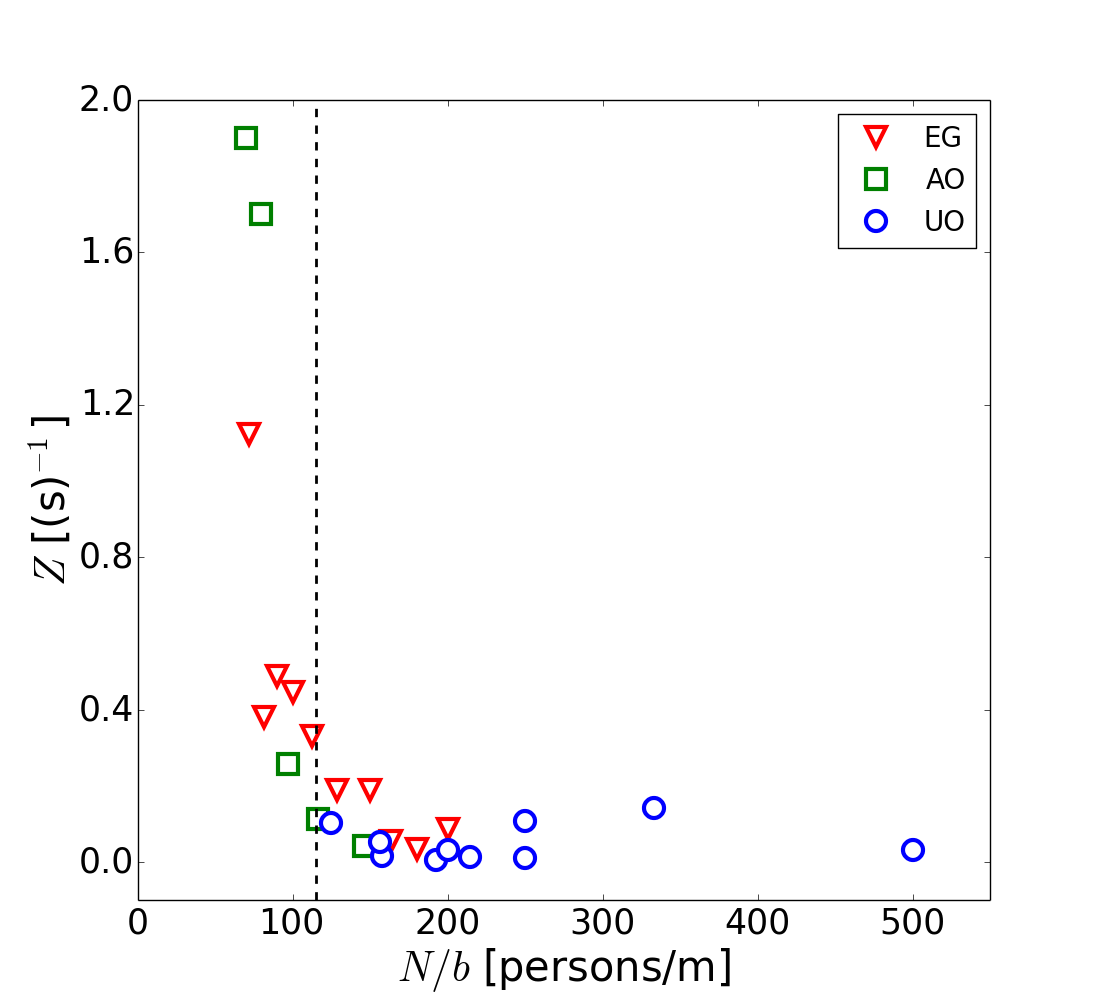}
    \caption{Relationship between the difference \emph{Z} and the ratio \emph{N/b}. \emph{Z} is calculated by using the flow in all states minus the flow in steady state and taking the absolute value. The black dashed line corresponds to \emph{N/b} = 115 persons/m, which is the critical value of the ratio \emph{N/b} to determine if the difference \emph{Z} is acceptable.}
\end{figure}

\section{Conclusions}

In this work, a feasible CUSUM algorithm is proposed to automatically detect steady state from density and speed time series of bottleneck experiments. To improve the algorithm, a step function is introduced to calculate the statistics for enhancing the sensitivity of the response to the fluctuations. In addition, a boundary is added to limit the increase of the statistics. The threshold of the detection parameter in the algorithm is calibrated using an autoregressive model. Comparing the detected steady states with previous manually selected ones, the modified CUSUM algorithm gives more reproducible results.

For the applications, three groups of bottleneck experiments are analysed by the modified CUSUM algorithm. The steady states of density and speed in each run are detected separately. Then the interval which is included both in the steady states of density and speed is regarded as the steady state of the flow. The flows in all states and in steady state are measured and compared. The results reconfirm that the flow per unit width is a constant as bottleneck width changes. Furthermore, the difference between the flows in all states and in steady state mainly depends on the ratio of pedestrian number to bottleneck width. The critical value of the ratio is approximately 115 persons/m. When the value of the ratio is higher than the critical value, the steady state is quickly reached and lasts longer. In this situation, the flow in all states is almost identical with the flow in steady state. Considering the ratio, we have more possibilities to compare the flows from different experiments, especially when the detection of steady state is difficult.

In future studies, the flow in steady state should be used when combing the flows from different bottleneck experiments. For the experiments in which it is difficult or impossible to detect steady state, the ratio of pedestrian number to bottleneck width should be calculated to estimate the difference between the flows in all states and in steady state.

\newpage

\section*{Appendix}

The matrix $M$ in Equation (13) has a size of $(K+1)(s_{\max}+1) \times (K+1)(s_{\max}+1)$. Solving $M$ directly with Gaussian elimination takes $O(s_{\max}^{3} K^{3})$ operations, which is not feasible for great $s_{\max}$.

Since $M$ is of block-tridiagonal structure, a Block Thomas-Algorithm \cite{Quarteroni2007} is used to solve the system efficiently in $O(s_{\max} K^{3})$. Essentially it is a block matrix variant of the Thomas-Algorithm, which is a simplified version of the Gaussian elimination on tridiagonal matrices.

The outline of the algorithm is as follows:

\[
M{\mathbf x}=
\begin{bmatrix}
   {D_1} & {U_1} & {   } & {   } & { 0 } \\
   {L_1} & {D_2} & {U_2} & {   } & {   } \\
   {   } & {L_2} & {D_3} & \ddots & {   } \\
   {   } & {   } & \ddots & \ddots & {U_{s_{\max}}} \\
   { 0 } & {   } & {   } & {L_{s_{\max}}} & {D_{s_{\max}+1}} \\
\end{bmatrix}\begin{bmatrix}
   {x_1} \\
   {x_2} \\
   { \vdots  } \\
   {x_{s_{\max}}} \\
   {x_{s_{\max}+1}} \\
\end{bmatrix}
=
\begin{bmatrix}
   {r_1} \\
   {r_2} \\
   { \vdots  } \\
   {r_{s_{\max}}} \\
   {r_{s_{\max}+1}} \\
\end{bmatrix},
 \]

where $D_i, U_i, L_i$ are matrices of size $(K+1) \times (K+1)$ and $x_i, r_i$ are vectors of size $K+1$. Since the zero blocks on the off diagonals are not necessary, using a sparse storage scheme reduces the memory amount for the matrix from $((K+1)(s_{\max}+1))^{2}$ to $(3s_{\max}+1) (K+1)^{2}$.

To obtain a solution, modify the entries in a forward sweep:
\begin{align*}
U_1^* &= D_1^{-1} U_1, \\
r_1^* &= D_1^{-1} r_1,
\end{align*}
for $i = 1$, and
\begin{align*}
U_i^* &= (D_i- L_i U_{i-1}^*)^{-1} U_i, \\
r_i^* &= (D_i- L_i U_{i-1}^*)^{-1} (r_i -L_i r_{i-1}^*),
\end{align*}
for $i = 2,3,\ldots,s_{\max}+1$.

Then in a backward sweep:
\begin{align*}
x_{s_{\max}+1} &= r^*_{s_{\max}+1},
\end{align*}
for $i = s_{\max}+1$, and
\begin{align*}
x_i &= r^*_i - U^*_i x_{i+1},
\end{align*}
for $i = s_{\max},s_{\max}-1,\ldots,2,1$.

\newpage

\bibliographystyle{unsrt}


\end{document}